\documentclass[authoryear,preprint,review,12pt]{elsarticle}
\usepackage{amssymb}
\usepackage{amsmath}
\usepackage{graphicx}
\begin{document}

\begin{frontmatter}

\title{Estimating financial risk using piecewise Gaussian processes}

\author[a]{I. Garc\'{i}a\fnref{*}}\ead{gmosquer@usb.ve} \author[b,c]{J. Jim\'{e}nez}

\address[a]{Departamento de C\'{o}mputo Cient\'{i}fico y Estad\'{i}stica
and Centro de Estad\'{i}stica y Software Matem\'{a}tico (CESMa),
Universidad Sim\'{o}n Bolivar, Sartenejas, Venezuela}

\address[b]{Laboratorio de Fen\'{o}menos no Lineales, Escuela de
F\'{i}sica, Facultad de Ciencias, Universidad Central de
Venezuela}
\address[c]{Red de Estudios Interdisciplinarios, Academia
Nacional de Ciencias F\'{i}sicas, Matem\'{a}ticas y Naturales,
Venezuela}

\fntext[*]{Telefax:+582129063364}
\begin{abstract}
We present a computational method for measuring financial risk by
estimating the Value at Risk and Expected Shortfall from financial
series. We have made two assumptions: First, that the predictive
distributions of the values of an asset are conditioned by
information on the way in which the variable evolves from similar
conditions, and secondly, that the underlying random processes can
be described using piecewise Gaussian processes. The performance of
the method was evaluated by using it to estimate $VaR$ and $ES$ for a
daily data series taken from the S\&P500 index and applying a
backtesting procedure recommended by the Basel Committee on
Banking Supervision. The results indicated a satisfactory
performance.
\end{abstract}


\end{frontmatter}

\section{Introduction}
Since the classic study by \cite{1} on the probability approach in econometrics, there has been a tendency to assume that a series of values of an asset, $\{V_{\tau};\ \tau=1,\ldots, t\}$, represents a random process with density $\rho_{\tau}(V)$. In fact, from the point of view of financial practice, what is interesting is to be able to make predictions conditioned by a particular information set. Thus, what we really want to model is the density $\rho_{t+1}(V|I_t)$, conditional on the information $I_t$ known at t, from which the expected price of the asset:
$$
\bar{V}_{t+1}=\int_{0}^{\infty} V\rho_{t+1}(V|I_t)dV
$$
and the deviations in these prices
$$
\sigma_{t+1}=\sqrt{\int_{0}^{\infty}(V-\bar{V}_{t+1})^{2}\rho_{t+1}(V|I_t)dV}
$$
can be calculated.

The expected return, $\bar{R}_{t+1}=(\bar{V}_{t+1}-V_t)/V_t$, can thus be obtained from the first integral, whilst the standard deviation is a measure of the uncertainty that is directly related to indexes commonly used in risk management, such as the Value at Risk ($VaR$) and Expected Shortfall ($ES$). The problem, of course, is that there is only one realization for each random process, thus the conditional distribution is an unobservable quantity that can only be inferred.

In this context, this study has a double aim: firstly, to use (in a way that we will describe in detail later) information about the evolution of prices from similar initial conditions, as the information $I_t$ that conditions the distribution we are looking for, and secondly, to design an inference model based on the assumption that the underlying random processes are piecewise Gaussian processes.

Due to the simplicity and versatility of Gaussian Processes ($GP$) for the modeling of arbitrary functions, they have become the basis for several techniques that have been developed to analyze several types of spatial and temporal problems (see e.g. \cite{2}, \cite{3}, \cite{4}, \cite{5}, \cite{13}, \cite{14}, \cite{6} and \cite{7}). In addition, modeling with $GP$ has an advantage that makes it particularly attractive from an econometric point of view, which is that it does not only permit predictions of the values of a series to be made, but also generates predictive distributions. This means that both the volatilities (given by the standard deviations of the predictive distributions) and indexes of risk such as $VaR$ can be estimated directly. Nevertheless, to the best of our knowledge, this approach has not been previously used for the estimation of financial risk indexes.

Furthermore, our $GP$ based models are local in the sense that the involved parameters depend on the information $I_t$, that in turn changes with time. As a consequence, the conditional predictive variances are time dependent and the models are thus heteroscedastic. In contrast to the ARCH/GARCH type models, however, there is no equation that controls the evolution of the volatility (see \cite{8}), and this change is instead driven by changes in $I_t$.

In the following section we introduce the specific type of piecewise models that we apply to estimate the risk in a financial time series. In section III we apply a validation procedure (backtesting) to evaluate the model according to recommendations given by the Basel Committee on Banking Supervision (\cite{10}, \cite{11} and \cite{9}). Concluding remarks are presented in Section IV.

\section{Prediction of the Value at Risk and Expected Shortfall
using Piecewise Gaussian Processes}
In what follows, we consider the simple case where we wish to make an estimation, $\hat{\rho}_{t+1}(V|I_t)$, of the conditional distribution $\rho_{t+1}(V|I_t)$ for a known series of the values of an asset $V_1, V_2, \ldots, V_t$, in order to predict the return one step ahead:
\begin{equation}\label{1}
  \hat{R}_{t+1}=(\hat{V}_{t+1}/V_t)-1;\ \hat{V}_{t+1}=\int_{0}^{\infty}
  V\hat{\rho}_{t+1}(V|I_t)dV
\end{equation}
and the volatility measured by the deviation:
\begin{equation}\label{2}
  \hat{\sigma}^R_{t+1}\equiv \frac{\hat{\sigma}_{t+1}}{V_t}=
  \frac{1}{V_t}\sqrt{\int_{0}^{\infty}(V-\hat{V}_{t+1})^{2}\hat{\rho}_{t+1}(V|I_t)dV}
\end{equation}

Specifying the information by which it is conditioned is therefore essential. We thus assume, as for the majority of the models used in financial econometrics, that events which occurred most recently are of greater relevance when we wish to establish conditions in the near future. This is the same as assuming that $I_t$ is contained in the sequence of values of the recent past: $V_{t-l+1},V_{t-l+2},\ldots,V_{t-1},V_t$, corresponding to a rolling window of size $l$. Part of the original contribution of our proposal has to do with the way in which $I_t$ is extracted from this data window. In this sense, our specific interest is focused on using the values of the asset which follows patterns of behavior of the data that are similar to those observed in the current window.

In order to explain in more detail what we mean, let's start by considering all possible segments of historical information that can be constructed with $l$ consecutive values of the prices of an asset:
$$
\begin{array}{ccc}
  \mathbf{v}_1 & = & (V_1,V_2,\ldots , V_l) \\
  \mathbf{v}_2 & = & (V_2,V_3,\ldots , V_{l+1}) \\
   & \vdots & \\
  \mathbf{v}_{t-l+1} & = & (V_{t-l+1},\ldots ,V_{t-1},V_t)
\end{array}
$$
The behavior patterns are understood to be the standardizations:
\begin{eqnarray}
  \nonumber \mathbf{zv}_{\tau}&=& (zV_{\tau},zV_{\tau+1},\ldots ,zV_{\tau +l-1});\
  \hspace{0.1cm} zV_{s}=\frac{V_s-\bar{V}_{\tau}}{\sigma_{V_{\tau}}}; \hspace{0.1cm}\
  \tau \leq s \leq \tau +l-1\\
  \hspace{2cm} \label{3}\\
  \nonumber \bar{V}_{\tau}&=& \frac{1}{l}\sum_{s=\tau}^{\tau +l-1}V_s; \hspace{0.1cm} \
  \sigma_{V_{\tau}}=\sqrt{\frac{1}{l-1}\sum_{s=\tau}^{\tau +l-1}(V_s-\bar{V}_{\tau})^{2}}
\end{eqnarray}
where $1\leq \tau \leq t-l+1$, such that even when the scales of the asset values may be very different between windows, the patterns are scale free profiles (see Figure 1).
\begin{figure}
  \includegraphics[width=12.0cm]{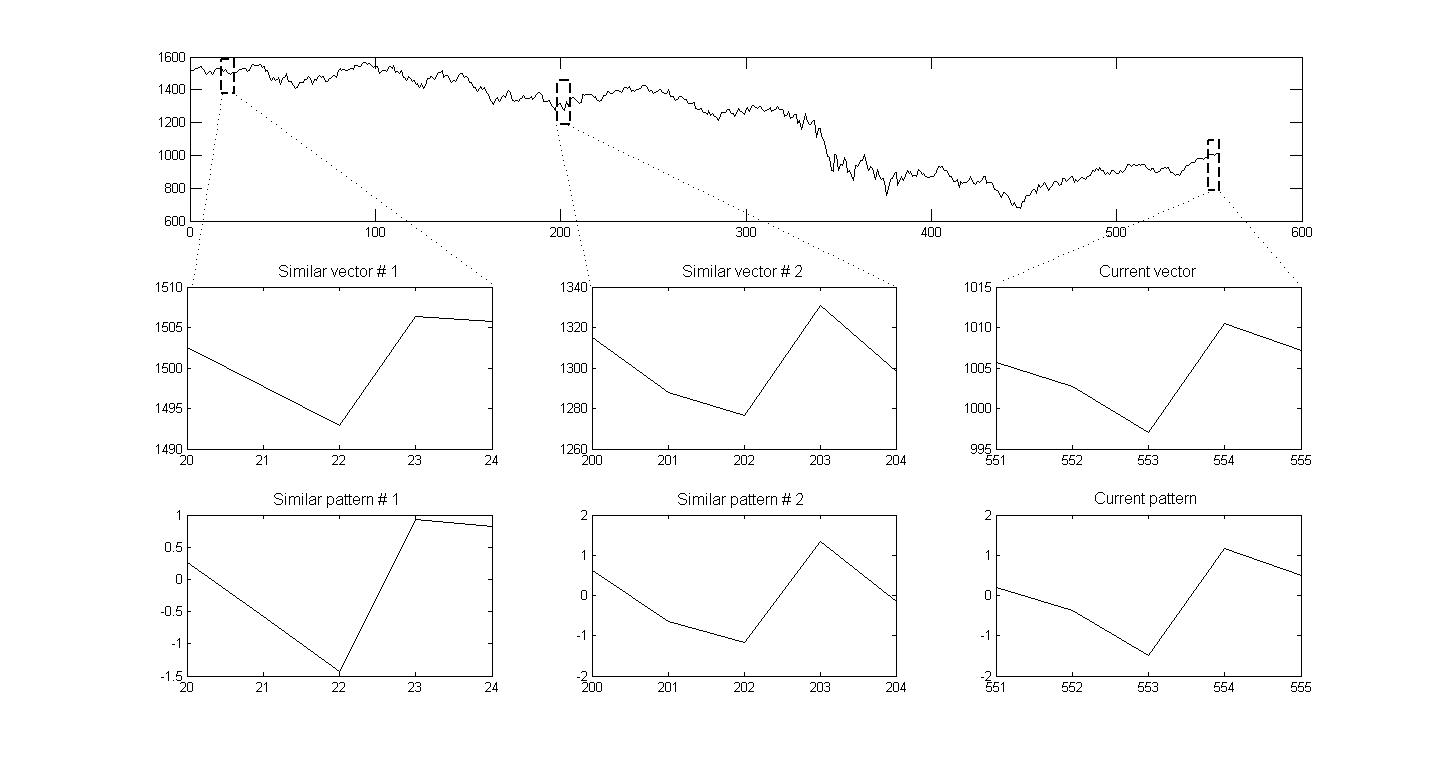}\\
  \caption{\emph{An example of what results
from the process of seeking patterns similar to the current
pattern. Standardization permits us to compare each past pattern
with the current pattern without taking into account the scale.
Thus, what we use as inputs for the local GP are similar
profiles.}}\label{figure 1}
\end{figure}

Let´s look now at the set of the $k$ $l$-dimensional behavior patterns that are most similar, as regards to their Euclidean distances, to the current pattern $\mathbf{zv}_{t-l+1}$. Thus we need to identify $k$ vectors that correspond to the smallest values of:
$$
   D_{t-l+1,\tau} = \| \mathbf{zv}_{t-l+1}-\mathbf{zv}_{\tau} \|
$$
and place them in increasing order:
$$
    \mathbf{zv}_{n_1(t,l)},\mathbf{zv}_{n_2(t,l)},\ldots ,\mathbf{zv}_{n_k(t,l)}
$$
where $ D_{t-l+1,n_1(t,l)} \leq D_{t-l+1,n_2(t,l)} \leq \cdots
\leq D_{t-l+1,n_k(t,l)}$.

Now we construct, by appropriately scaling each $V_{n_i(t,l)+l}$ observed one step ahead of the i-th pattern, the respective standardized values of the asset which correspond to each of the patterns that define the neighborhood of the current pattern:
\begin{equation}\label{5}
    \mathbf{zv}_{n_i(t,l)}\rightarrow zV_{n_i(t,l)+l}\equiv
    \frac{V_{n_i(t,l)+l}-\bar{V}_{n_i(t,l)}}{\sigma_{V_{n_i(t,l)}}};\
    1\leq i\leq k
\end{equation}

What we assume as being the information $I_t$, is the set of ordered pairs:
\begin{equation}\label{6} \mathbf{X}(t,l)\equiv
\{\mathbf{x}_i(t,l),y_i(t,l);\ 1\leq i \leq k\};
\end{equation}
where
\begin{equation}\label{7} \mathbf{x}_i(t,l)\equiv
\mathbf{zv}_{n_i(t,l)},\ y_i(t,l)\equiv zV_{n_i(t,l)+l}
\end{equation}

The second step in our method is to adjust the hyperparameters of the $GP$ represented by a set of random variables $f_t(\mathbf{x}_i(t,l))$ and a noise term $\epsilon_i$ that satisfy the local condition:
\begin{equation}\label{8}
 y_i(t,l)=f_t(\mathbf{x}_i(t,l))+\epsilon_i;\ \
 \epsilon_i\ \stackrel{\mathrm{iid}}{\sim}\ \mathcal{N}(0,\nu^2_{t}),\ \
 i=1,2,\ldots,k
\end{equation}
which we then use to generate the conditional predictive
distribution
$$g(y^*(t,l)|\mathbf{X}(t,l),\mathbf{x}^*(t,l))$$
where $\mathbf{x}^*(t,l)\equiv \mathbf{zv}_{t-l+1}$.

By following the standard procedure used in these cases and that is described in the relevant literature (see e.g. \cite{2}, \cite{14} and \cite{12}), we obtain that:
\begin{equation}\label{9}
    g(y^*(t,l)|\mathbf{X}(t,l),\mathbf{x}^*(t,l))
    \propto exp\left[-\frac{(y^*(t,l)-\bar{y}(t,l))^2}{2\hat{\sigma}^{*2}(t,l)}\right],
\end{equation}
where $\bar{y}(t,l)$ can be identified as the expected value of $y^*(t,l)$ and $\hat{\sigma}^{*2}(t,l)$ with the variance, which are given respectively by:
\begin{equation}\label{10} \bar{y}(t,l)=
\mathbf{c}(t,l)\mathbf{C}^{-1}(t,l)\mathbf{y}(t,l);\
\hat{\sigma}^{*2}(t,l)=\gamma(t,l)-\mathbf{c}(t,l)\mathbf{C}^{-1}(t,l)\mathbf{c}^T(t,l)
\end{equation}
with:
\begin{equation}\label{11}
\mathbf{c}(t,l)=(c(\mathbf{x}^*(t,l),\mathbf{x}_1(t,l)),c(\mathbf{x}^*(t,l),\mathbf{x}_2(t,l)),\ldots
,c(\mathbf{x}^*(t,l),\mathbf{x}_k(t,l))),
\end{equation}
$\mathbf{C}^{-1}(t,l)$ is the inverse matrix of:
\begin{equation}\label{12}
\left(%
\begin{array}{cccc}
  c(\mathbf{x}_1(t,l),\mathbf{x}_1(t,l)) & c(\mathbf{x}_1(t,l),\mathbf{x}_2(t,l)) & \ldots & c(\mathbf{x}_1(t,l),\mathbf{x}_k(t,l)) \\
  c(\mathbf{x}_2(t,l),\mathbf{x}_1(t,l)) & c(\mathbf{x}_2(t,l),\mathbf{x}_2(t,l)) & \ldots & c(\mathbf{x}_2(t,l),\mathbf{x}_k(t,l)) \\
  \vdots & \vdots & \ddots & \vdots \\
  c(\mathbf{x}_k(t,l),\mathbf{x}_1(t,l)) & c(\mathbf{x}_k(t,l),\mathbf{x}_2(t,l)) & \ldots& c(\mathbf{x}_k(t,l),\mathbf{x}_k(t,l)) \\
\end{array}%
\right)
\end{equation}
\begin{equation}\label{13}
\mathbf{y}(t,l)=\left(%
\begin{array}{c}
  y_1(t,l) \\
  y_2(t,l) \\
  \vdots \\
  y_k(t,l) \\
\end{array}%
\right);\ \gamma(t,l)=c(\mathbf{x}^*(t,l),\mathbf{x}^*(t,l))
\end{equation}
the superindex $^T$ denotes transposition, and $c(\mathbf{a},\mathbf{b})$ is the covariance function defining the $GP$, here assumed as:
\begin{equation}\label{14}
c(\mathbf{a},\mathbf{b})=\theta^2_t\exp{\left[-\frac{1}{2}(\mathbf{a}-\mathbf{b})^T
 M_t(\mathbf{a}-\mathbf{b})\right]}
\end{equation}
where $M_t$ is the diagonal matrix:
$$
\left(%
\begin{array}{ccc}
  1/s_{1t}^2 & \ldots & 0 \\
  \vdots & \ddots & 0  \\
  0 & 0 & 1/s_{dt}^2 .\\
\end{array}
\right)%
$$
In order to predict the asset value $\hat{V}_{t+1}$, we take the inverse of the standardization process on the expected value $\bar{y}(t,l)$:
\begin{equation}\label{15}
\hat{V}_{t+1}=\sigma_{V_{t-l+1}}\cdot\bar{y}(t,l)+\bar{V}_{t-l+1},
\end{equation}
while the prediction for the uncertainty $\hat{\sigma}_{t+1}$ is:
\begin{equation}\label{16}
 \hat{\sigma}_{t+1}=\sigma_{V_{t-l+1}}\cdot\hat{\sigma}^{*}
\end{equation}

It is worth emphasizing the local character of the proposed scheme, in the sense that in order to obtain the predictive distribution at $t + 1$ we use a $GP$ whose hyperparameters $(\nu^2_{t},\theta^2_t,s_{1t},s_{2t},\ldots,s_{lt})$ are fitted to the information $I_t$. This itself is dependent on time, as indicated by Eq. \ref{3}-\ref{7}, causing the $GP$ parameters to change, providing a sufficiently flexible method for the modeling of non stationary series. As regards to the manner in which the
parameters that intervene in the co-variance matrix are fixed, a criteria that is frequently used is to maximize the log-likelihood:
\begin{equation}\label{17}
    \mathcal{L}=-\frac{1}{2}log[det(\mathbf{C}(t,l))]-\frac{1}{2}\mathbf{y}(t,l)^T\mathbf{C}^{-1}(t,l)\mathbf{y}(t,l)
    -\frac{t}{2}log2\pi .
\end{equation}
Furthermore, as indicated by the result of Eq. \ref{9}, when we apply models based on $GP$ we obtain a predictive distribution, which makes it relatively simple to estimate the probability that a given event will occur. This fact is extremely interesting, from the financial point of view, in decision making. In particular, we show how models based on $GP$ permit a straightforward evaluation of both $VaR$ and $ES$.

We start by calculating the Value at Risk for a time horizon $T$ and probability $\alpha \in (0,1)$. The usual definition is:
\begin{equation}\label{18}
VaR_{t+T}^{\alpha}=sup\{v\ ; \ P[(R_{t+T}\leq v)|I_t]\leq \alpha
\}
\end{equation}
where $P(R|I_t)$ is the conditional distribution of the probabilities of the asset returns and $I_t$ is, as before, the information that conditions what occurs at $t + T$. In other words, $VaR$ is a simple way to estimate the minimal potential loss that, for a given time horizon, will be contained in the $100\alpha$\% of the worst case scenarios.

$VaR$ may be estimated directly with the method we have described: we only have to calculate the return corresponding to the value, $\hat{V}_{t+1}^{\alpha}$, below which the price of the asset may be found with probability $\alpha$. The predictive distribution of the prices is given by the $GP$, thus $\hat{V}_{t+1}^{\alpha}$ satisfies the condition:
\begin{equation}\label{19}
    \alpha=\frac{1}{Z}\int_0^{\hat{V}_{t+1}^{\alpha}}
    \exp\left[- \frac{(x-\hat{V}_{t+1})^2}{2\hat{\sigma}_{t+1}^2}\right]dx
\end{equation}
where $Z$ is the normalization factor:
\begin{equation}\label{20}
    Z=\int_0^{\infty}\exp\left[ -\frac{(x-\hat{V}_{t+1})^2}{2\hat{\sigma}_{t+1}^2}\right]dx=
    \sqrt{\frac{\pi}{2}}\hat{\sigma}_{t+1}\left[ 1+erf\left(\frac{\hat{V}_{t+1}}{\sqrt{2}\hat{\sigma}_{t+1}}\right)\right]
\end{equation}
and $erf(x)$ is the error function:
\begin{equation*}
    erf(x)=\frac{2}{\sqrt{\pi}}\int_0^x \exp{(-\,\,x^2)}dx
\end{equation*}
Combining Eq. \ref{19} and Eq. \ref{20} we obtain:
$$
\sqrt{\frac{\pi}{2}}\,\,\,\hat{\sigma}_{t+1}\left[erf\left(\frac{\hat{V}_{t+1}}{\sqrt{2}\hat{\sigma}_{t+1}}\right)-
erf\left(\frac{\hat{V}_{t+1}-\hat{V}_{t+1}^\alpha}{\sqrt{2}\hat{\sigma}_{t+1}}\right)\right]=\alpha
Z=
$$
$$
 = \alpha\sqrt{\frac{\pi}{2}}\hat{\sigma}_{t+1}\left[ 1+
erf\left(\frac{\hat{V}_{t+1}}{\sqrt{2}\hat{\sigma}_{t+1}}\right)\right]
$$
and therefore
$$
erf\left(\frac{\hat{V}_{t+1}-\hat{V}_{t+1}^{\alpha}}{\sqrt{2}\hat{\sigma}_{t+1}}\right)=
(1-\alpha)erf\left(\frac{\hat{V}_{t+1}}{\sqrt{2}\hat{\sigma}_{t+1}}\right)-\alpha
$$
or
\begin{equation}\label{21}
\hat{V}_{t+1}^{\alpha}=\hat{V}_{t+1}-\sqrt{2}\hat{\sigma}_{t+1}erf^{-1}(\lambda_{t+1})
\end{equation}
with
\begin{equation}\label{22}
    \lambda_{t+1}=(1-\alpha)erf\left(\frac{\hat{V}_{t+1}}{\sqrt{2}\hat{\sigma}_{t+1}}\right)-\alpha
\end{equation}

According to Eq. \ref{21}, our estimation of $VaR_{t+1}^{\alpha}$ will
be given by:
\begin{equation}\label{23}
 \widehat{VaR}_{t+1}^{\alpha}= \hat{R}_{t+1}-\frac{\sqrt{2}\hat{\sigma}_{t+1}}{V_t}erf^{-1}(\lambda_{t+1})
\end{equation}
where $\hat{R}_{t+1}$ and $\hat{\sigma}_{t+1}$ are evaluated using Eq. \ref{1}, \ref{15} and \ref{16}, and $\lambda_{t+1}$
is given by Eq. \ref{22}.

We will now estimate $ES$ with our model. If $\rho_{t+1}(R|I_t)$ is the conditional distribution of the returns of an asset and $VaR_{t+1}^{\alpha}$ the respective value at risk for a level $\alpha$, then the expected shortfall, $ES$, is:
\begin{equation}\label{24}
    ES_{t+1}^{\alpha}=E(R_{t+1}|I_t,(VaR_{t+T}^{\alpha}>R_{t+1}))\propto
\int_{-\infty}^{VaR_{t+1}^{\alpha}}R\rho_{t+1}(R|I_t)dR
\end{equation}
such that, while $VaR$ gives information about the least loss we could expect at a given confidence level; $ES$ informs the worse expected loss that we can expect to occur.

In those cases where hedging needs to be designed, an adequate estimate of the $ES$ is more informative than $VaR$. Let´s take as a hypothetical situation that at a particular time $t$ we know that the $VaR_{t+1}^{0.01}$ of an asset or a portfolio is $-0.001$. There is thus a 99\% probability that the return will exceed this
value. However, the hedging plan will change if we estimate that the expected return for the remaining 1\% is either $-0.002$ or
$-0.1$.

As for $VaR$, it is relatively simple to estimate $ES$ using our $GP$ based model. Effectively, since the predictive distributions of the prices are Gaussian, so are those of the arithmetic returns (given that these are obtained from the prices by dividing them by yesterday´s price). Thus, we can evaluate the previous integral by
calculating the expected value $\widehat{ESV}_{t+1}^{\alpha}$ of the price of the asset, as long as this is below the $100\alpha$\% quantile of the predictive distribution of the prices
$\hat{V}_{t+1}^{\alpha}$
\begin{equation}\label{25}
    \widehat{ESV}_{t+1}^{\alpha}\equiv
\frac{1}{C_{t+1}^{\alpha}}\int_0^{\hat{V}_{t+1}^{\alpha}}x\exp\left[-\frac{(x-\hat{V}_{t+1})^2}
{2\hat{\sigma}_{t+1}^2}\right]dx,
\end{equation}
where $C_{t+1}^{\alpha}$ is the normalization factor:
\begin{eqnarray*}
 C_{t+1}^{\alpha} &=& \int_0^{\hat{V}_{t+1}^{\alpha}}
\exp\left[-\frac{(x-\hat{V}_{t+1})^2}{2\hat{\sigma}_{t+1}^2}\right]dx= \\
   &=& \sqrt{\frac{\pi}{2}}\,\,\,\hat{\sigma}_{t+1}\left[erf\left(\frac{\hat{V}_{t+1}}{\sqrt{2}\hat{\sigma}_{t+1}}
\right)-erf\left(\frac{\hat{V}_{t+1}-\hat{V}_{t+1}^\alpha}{\sqrt{2}\hat{\sigma}_{t+1}}\right)\right],
\end{eqnarray*}

The integral in Eq. \ref{25} can be readily evaluated:
$$
    \widehat{ESV}_{t+1}^{\alpha}= \hat{V}_{t+1}- \sqrt{\frac{2}{\pi}} \,\,
    \hat{\sigma}_{t+1} \left\{ \frac{\left[\exp\left(- \,\,\frac{(\hat{V}_{t+1}^\alpha -
    \hat{V}_{t+1})^2}{2\,\,\hat{\sigma}_{t+1}^2}\right)-\exp\left(-\,\,\frac{\hat{V}_{t+1}^2}{2\,\,
    \hat{\sigma}_{t+1}^2}\right)\right]}{\left[erf\left(\frac{\hat{V}_{t+1}}{\sqrt{2}\,\,
    \hat{\sigma}_{t+1}}\right)-erf\left(\frac{\hat{V}_{t+1}-\hat{V}_{t+1}^\alpha}{\sqrt{2}\,\
    ,\hat{\sigma}_{t+1}}\right)\right]}\right\}
$$
and from this result we get our estimation for $\widehat{ES}_{t+1}^{\alpha}$:
\begin{eqnarray}
\nonumber \widehat{ES}_{t+1}^{\alpha}&=&\frac{\widehat{ESV}_{t+1}^{\alpha}}{V_t}-1=\hat{R}_{t+1}-\\
   &-& \sqrt{\frac{2}{\pi}} \,\,
    \frac{\hat{\sigma}_{t+1}}{V_t} \left\{ \frac{\left[\exp\left(- \,\,\frac{(\hat{V}_{t+1}^\alpha -
    \hat{V}_{t+1})^2}{2\,\,\hat{\sigma}_{t+1}^2}\right)-\exp\left(-\,\,\frac{\hat{V}_{t+1}^2}{2\,\,
    \hat{\sigma}_{t+1}^2}\right)\right]}{\left[erf\left(\frac{\hat{V}_{t+1}}{\sqrt{2}\,\,
    \hat{\sigma}_{t+1}}\right)-erf\left(\frac{\hat{V}_{t+1}-\hat{V}_{t+1}^\alpha}{\sqrt{2}\,\
    ,\hat{\sigma}_{t+1}}\right)\right]}\right\}\label{26}
\end{eqnarray}
In the next section we evaluate the performance of the model by showing the results obtained by calculating $\widehat{VaR}_{t+1}^{0.01}$ following Eq. \ref{23} and $\widehat{ES}_{t+1}^{0.01}$ according to Eq. \ref{26}, for a series taken from the S\&P500 index.

\section{Backtesting VaR and ES estimations}
The set of recommendations established by the Basel Committee is currently used to measure the quality of $VaR$ estimates. These are statistical validation measurements (backtesting) based on the quantification of exceptions:
$$
e_{t+1}^{\alpha}=\Theta(\widehat{VaR}_{t+1}^{\alpha}-R_{t+1})
$$
that can occur when estimating $VaR_{t+1}^{\alpha}$, where $\Theta(x)$ is Heaviside´s step function ($\Theta(x)=1(0)$ when $x > 0$ (in any other case)).

Whichever the method used for calculating $VaR$, the condition that must be satisfied is:
$$
E(e_{t+1}^{\alpha}|I_t)=\alpha
$$
and what is examined are the different consequences of this condition.

Probably, the most utilized format is that if there are a total of $n$ calculated values, the total number of exceptions:
$$
x=\sum_{t=0}^{n-1} e_{t+1}^{\alpha}
$$
follows a binomial distribution with parameters $n$ and $\alpha$. Thus for a 5\% confidence level for $n = 250$ evaluations (that for daily data represents approximately one year) and a value of $\alpha=0.01$, the method will only be rejected when $x > 5$.

For the example of the series taken from the S\&P500 index represented in Figure 2, we can use Eq. \ref{23} to calculate $\widehat{Var}_{t+1}^{0.01}$ over three consecutive windows of $n = 250$ data points (i.e. approximately three continuous years) and not once do we obtain more exceptions than those considered acceptable according to the validation criteria (3, 5 and 5 exceptions, respectively).

\begin{figure}
  \includegraphics[width=12.0cm]{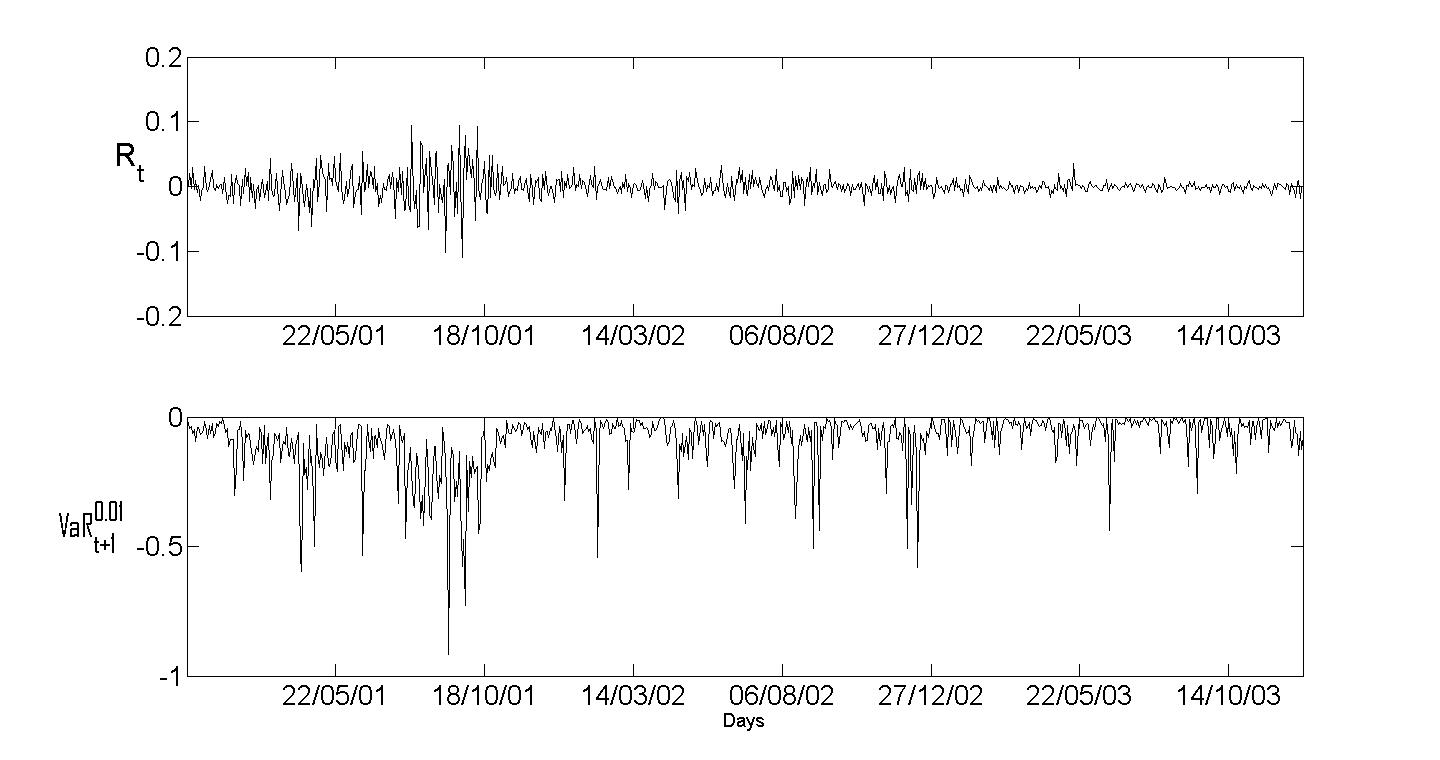}\\
  \caption{\emph{In the upper part of the figure a segment representing
the daily returns of the S\&P500 index over a period of three
years is shown. Observe the period of high volatility
characterized by large fluctuations in the returns. The inferior
part of the figure shows our $VaR_{t+1}^{0.01}$
estimates}.}\label{figure 2}
\end{figure}

As regards the evaluation of $ES$ estimates done with the model, for each of the exceptions (total of 13) we compared the $ES$ that we get using our model with the observed returns. Specifically, if we use $t_i^e, i=1,\ldots, 13$ to denote the 13 values of the times that correspond to the exceptions, we can calculate their mean squared error:
$$
\frac{1}{\rho_e}\sqrt{\sum_{i=1}^{13}\left(\widehat{ES}_{t_i^e}-
R_{t_i^e}\right)^2},\ \rho_e=
\sqrt{\sum_{i=1}^{13}\left(R_{t_i^e}-\frac{1}{13}\sum_{j=1}^{13}R_{t_j^e}
\right)^2}
$$
This gives an error of 0.12, indicating the quality of the fit
achieved by our model.

\section{Discussion}
There are several reasons why the behavior of financial markets is difficult to model, above all, because each agent involved prefers that others do not have access to a model that permits them to correctly anticipate fluctuations in the market. If they did, and could thus exchange uncertainty for predictability, circumstantial informational advantages that could be used to increase net worth would be lost, which at the end of the day is what this is all about.

Apart from practical motivations whose importance is obvious, the difficulties inherent to modeling combined with the ever increasing possibility of having access to large volumes of data and computer software, have stimulated the development of empirical models that, as well as predicting what the system will or will not do, also attempt to estimate the degree of uncertainty that this may present.

Models based on Gaussian processes incorporate this option in a direct way as a consequence of their structure. This is advantageous due to the simplification of the theory involved and also from a practical point of view. They do this by assuming that the series of values of an asset is a realization of a random process and that this is Gaussian. Thus, predictive conditional distributions that are Gaussian centered on the expected values of the asset are obtained, whose standard deviations fix the respective volatility values and whose quantiles correspond to estimates of risk indexes such as $VaR$ and $ES$. All this without the necessity of introducing additional assumptions, such as for example, the evolution of the conditional variance and/or the distribution of the observed returns, which are necessary in ARCH/GARCH type models and in models of stochastic volatility.

Another beneficial aspect is the ease with which we can include the assumptions as relevant information that condition the predictive distribution in each case. One of these benefits can be when we suppose that the information is given by the way in which the system evolves from similar initial situations. The problem is thus reduced to specifying the set of vectors that this information represents and to evaluate the corresponding covariance function.

With regards this last point, it is woth mentioning that even though all the results we have presented correspond to a particular selection of the covariance function, the key issue is that whatever the proposed function is, this generates a covariance matrix whose elements are much higher among vectors that are similar than among those that are not, which falls rapidly with the distance between vectors. Finally, the models presented here permit the establishment of simple criteria for risk evaluation that could be useful in the design of dynamic portfolios with optimal risk, which could in turn result in portfolios that generate higher equity than that produced by minimum risk portfolios. These are two of the aspects we wish to focus on in a further study.

\section{Acknowledgement}
This paper is part of the PhD Thesis of I. Garc\'{i}a at Universidad Sim\'{o}n Bol\'{i}var, which was financed through a fellowship by the Academia Nacional de Ciencias F\'{i}sicas, Matem\'{a}ticas y Naturales of Venezuela.
Venezuela.

\end{document}